# Flash-based Audio and Video Communication in the Cloud

Kundan Singh
Intencity Cloud Technologies,
San Francisco, CA, USA

kundan10@gmail.com

Carol Davids
Illinois Institute of Technology,
Wheaton, IL, USA

davids@iit.edu

## ABSTRACT
Internet telephony and multimedia communication protocols have matured over the last fifteen years. Recently, the web is evolving as a popular platform for everything we do on the Internet including email, text chat, voice calls, discussions, enterprise apps and multi-party collaboration. Unfortunately, there is a disconnect between web and traditional Internet telephony protocols as they have ignored the constraints and requirements of each other. Consequently, the Flash Player is being used as a web browser plugin by many developers for web-based voice and video calls.

We describe the challenges of video communication using a web browser, present a simple API using a Flash Player application, show how it supports wide range of web communication scenarios in the cloud, and describe how it can interoperate with Session Initiation Protocol (SIP)-based systems. We describe both the advantages and challenges of Flash Player based communication applications. The presented API could guide future work on communication-related web protocol extensions.

## Categories and Subject Descriptors
D.2.11 [**Software Architectures**]: Web-based multimedia communication architecture

## General Terms
Design, Experimentation

## Keywords
Flash Player; Web video communication; cloud communication; SIP; RTMP; video conference

## 1. INTRODUCTION
The World-Wide-Web is powered by the Hypertext Transfer Protocol (HTTP), and languages such as the Hypertext Markup Language (HTML) and Javascript [1, 2]. Web protocol and languages work within constraints such as cross domain and device access restrictions. Many network firewalls open outbound TCP port 80 and 443 to allow client-server connections from a web browser to a server. On the other hand, Internet telephony and multimedia communication are typically enabled by the Session Initiation Protocol (SIP) and the Real-time Transport Protocol (RTP) [3, 4]. SIP-based Internet telephony has matured over the last decade and a half with many audio and video telephony products and services available today. A SIP user agent (UA) contains both a client and a server, to initiate and receive connections or requests. User agents communicate via centralized rendezvous servers to discover each other, and exchange in-call media packets end-to-end using RTP over UDP as shown in Fig.1. Traditionally, these protocols have ignored the constraints of web browsers such as missing a UDP transport and listening socket, and due to the restrictions of firewalls and network address translation (NAT), are difficult to implement within a web browser. Lack of SIP tools in the browser is a huge missed opportunity for communication with big impact on innovation. Unlike the few hundred organizations working on SIP there are millions of web developers who are holding the keys to the countless web applications and to innovation.

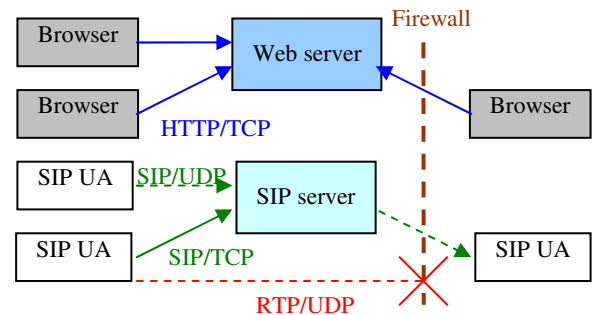

**Figure 1. Web vs SIP communication path**

With the popularity of the web, many applications are being moved to web servers and the cloud. The web browser has become the most important client application needed by end users. Cloud infrastructure has allowed organizations to build businesses solely on the web platform, e.g., email access, file backup storage, document editing and sharing, watching TV and movies, text chat, enterprise apps and multiparty collaboration. Hence, there is a huge demand to fill the gap caused by the disconnection between traditional Internet telephony protocols and web-based systems. This is evident from the new working groups' effort [6, 7] within W3C and IETF proposed by major industry leaders such as Google and Skype. However, like any other extension to web protocol and language, it will take time or may not happen until all major browsers incorporate the new standard and build a truly interoperable web-based communication platform. Moreover, some businesses prefer proprietary technology to standards for digital rights management of content or the walled garden nature of communication service providers.

Browser plugins such as Java applet and Flash Player have been used for many years to do web-based real-time communication. The Flash Player is available to almost everyone with an Internet connection, provides immersive communication experience within web browsing to end users, and is easy for developers to build on. Hence, many organizations are using Flash Player for real-time video communication and web-to-phone calls [12, 13]. With new extensions in the Flash Player for end-to-end media path and group communication, one can build a reusable web application gadget (widget) to enable wide range of web-based audio/video communication scenarios. Our Flash VideoIO project demonstrates such a widget and presents an easy to use API that enables web developers with only HTML and Javascript knowledge to build web-based video communication applications [10].

Unlike other approaches that separate the device, display and communication elements, we describe a simple browser plugin-based integrated component to make it easy for web developers to build applications. Using a simple Javascript API to set one src property of VideoIO instances, we demonstrate several different video communication scenarios such as a two-party call, multiparty conference, video messaging, broadcast of live event and panel discussion, white-labeled Internet phone and web video phone on Facebook's social network platform.

We will use here the term Flash Player in the understanding that it is more than just an audio/video media player, since it also executes application code. It provides secure access to the computer resources such as audio/video devices and implements hardware acceleration for most popular operating systems and processor types. It also supports UDP-based media transport and echo cancellation.

The paper is organized as follows. Section 2 lists related work in web and multimedia communications. Section 3 describes the challenges of web-based communications, advantages and issues of the Flash Player. We give the background of the Flash Player for video communications in Section 4. We present our Flash VideoIO API in Section 5 and describe real implementations of various web and cloud-based video communication scenarios in Section 6. Section 7 describes interoperability with SIP/RTP-based systems. Section 8 presents conclusions and future work.

## 2. RELATED WORK

Video communications are not new. As early as the 1960's there were video phones such as the picture phone. Today there are many video phones to choose from. Most of the early video phones carried encoded video bits over switched circuits, e.g., room-based H.320 video conferencing systems over ISDN and consumer H.324 video phones over phone modems [9].

With the popularity of Internet Protocol and IP multicast, people started building desktop video conferencing over the Internet2's multicast backbone (Mbone) using robust audio tool (rat) and video conferencing tool (vic) [14]. Subsequently, other related Internet applications such as voice-over-IP (VoIP) and video streaming became popular. Video became a natural extension to SIP and H.323-based VoIP systems. Streaming was amended to accommodate user interactivity for real-time video conferencing. In particular, Flash Player emerged as a popular browser plugin for video content. When Flash Player added interactivity using the Real Time Messaging Protocol (RTMP) [15], two-types of web applications appeared (Fig. 2): (a) web-to-phone call translated to SIP/RTP using a backend gateway such as Flash Media Gateway or our siprtmp, and (b) multiparty video conferences used backend interactive media streaming server.

A particular application inherits some advantages and problems based on its origin. For example, the H.323 applications inspired by video-over-phone were complex, and required large telecom investment to work with. IP multicast conferences used text-based protocols but could not be used on public Internet where multicast is unavailable. VoIP protocols largely ignored the constraints of web browser and web browsers ignored the transport requirements of VoIP protocols, hence these protocols are difficult to use on the web. Video streaming applications opened up to web platforms but inherited centralized client-server nature of media streaming not always suitable for real-time interactive communication.

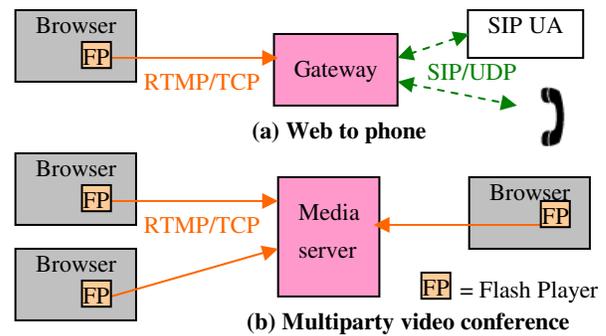

Figure 2. Flash-based communication

Web-to-phone systems have been around for a long time, e.g., Dialpad in 1999. In such systems, typically, a front-end Java applet or plugin application in the browser connects to a backend gateway to existing VoIP protocols such as SIP as shown in Fig. 2(a). Due to browser and network restrictions, the client-gateway transport is usually TCP, not UDP, which hurts the end-to-end media latency. The applet or plugin captures and plays real-time media because of the lack of capabilities in a browser. To solve the media latency problem, Flash Player enabled end-to-end media path over UDP using Real Time Media Flow Protocol (RTMFP) and group communication using application level multicast to bridge the web and communication systems [16].

New browser capabilities are being defined in HTML5 for video conferencing and peer-to-peer communication [5]. New working groups are forming in W3C and IETF to define elements of real-time communication in the browser [6, 7, 8]. HTML5 extensions, such as websocket, enable the asynchronous communication needed for web conference signaling. The main advantage of the approach is that it has no external dependencies on plugins or applications besides a web browser. However, (1) a change in the web protocol or language requires time before all major browsers have consistent implementations, (2) some businesses prefer a proprietary plugin over standard-based approach for digital rights management to protect their content or customer interactions, and (3) the security implications for end users is unclear if a web page can start end-to-end video calls between two users, or between users on two different web sites.

Researchers have proposed replacing SIP with HTTP on the web and using XML for session negotiation and Host Identity Protocol (HIP) for firewall and NAT traversal [17]. Any telephony function not related to the web is placed in external SIP gateways that translate between the web and SIP clouds for interoperability. The Voice and Video over Web project at the Illinois Institute of Technology is developing technology and sample implementations for web applications that include real-time voice communications [19]. The project aims for communication tools to become as common on web pages as other components such as layout, buttons, images and multimedia players. It proposes an external host resident application to extend the browser for communication using existing/new standards, unlike ours that relies on the proprietary protocols of Flash Player. The RESTful signaling mechanism used in that project can also be applied to other web communication projects such as our Flash-based application. In fact the preliminary implementation of [19] uses our Flash-based application to handle the audio and video capture, playback and transport.

# 3. CHALLENGES IN THE WEB PLATFORM

The web platform poses restrictions to video communication.

## 3.1 Browser Restrictions

A browser typically puts external web pages and applications in a secure and restricted sandbox. The cross domain restriction disallows a Javascript application downloaded from one domain to access data on another domain, or to make Ajax (Asynchronous Javascript and XML) request to another domain. The browser restricts access to host resources such as camera and microphone which prevents building a VoIP phone in HTML/Javascript.

HTTP is the only protocol between a browser and server because web languages do not currently support other application level protocols. Private networks usually prevent non-HTTP traffic using firewalls. Hence, application developers fall back to HTTP to make it work. The Flash Player's streaming protocol has an HTTP tunneling mode to bypass firewall restrictions.

Unlike one way streaming that can tolerate the network jitter and delay of TCP, bidirectional interactive real-time communication requires UDP for a low-latency end-to-end media path. Moreover, a web page cannot create a listening socket to receive connections from another browser instance. VoIP protocols such as SIP and RTP typically require support of UDP and/or listening sockets, and hence cannot be implemented in HTML/Javascript alone.

## 3.2 Plugin Restrictions

Some of these browser restrictions can be avoided by using browser plugins such as Java applet and Flash Player. However, these plugins have their own restrictions. For example, a Java applet can connect using TCP only to the same server hosting the applet.

A Flash application can create a general purpose outbound TCP socket to implement custom protocols as well as capture camera and microphone data from the host computer. This allows, in theory, the implementation of a SIP phone using TCP in an application. However, the Flash Player imposes cross domain restrictions for security reasons, where an application downloaded from domain A can connect to domain B only if domain B has given explicit permission to domain A. Hence, such a web phone cannot freely connect to other existing SIP servers in different domains, e.g., iptel.org.

Secondly, the Flash Player sandbox does not give access to *encoded* media data captured from a camera or microphone to the application. The only operation allowed is to attach the device to a network stream to send media to external media server as shown in Fig.3. Hence, a Flash application cannot implement custom protocols such as RTP for media transport.

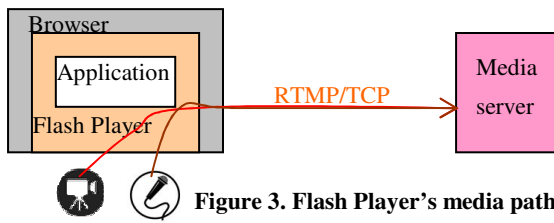

**Figure 3. Flash Player's media path**

A Flash application cannot create a custom UDP or listening socket, which prevents an end-to-end path in a communication application. However, it can use end-to-end proprietary RTMFP [16] if interoperability to standards-based system is not required.

The limited set of audio video codecs in the Flash Player also may not be compatible with existing SIP devices. Speex is the only supported open source audio codec. Even though it can play H.264 video, the camera capture data is restricted to a proprietary video encoder and hence cannot easily interoperate with existing video phones.

The audio and video quality is limited by what the plugin provides. The camera capture video encoder quality is not good enough for large displays. The lack of echo cancellation in the Flash Player had been a problem for web phone applications. Web sites recommend using a head-set, implement push-to-talk or an additional audio engine plugin, or perform network-based echo cancellation. Fortunately, the new version of Flash Player will have built-in acoustic echo cancellation.

Finally, even though Flash Player is available on major platforms and browsers, it is not available on all, e.g., iPhone. The closed nature and the business competition among organizations may cause it to never be available for such devices. With the growing number of custom smart devices, this becomes an issue for web developers if their applications cannot run on all platforms.

In general, a plugin imposes restrictions because of its closed nature but allows portability of application across different browsers because the plugin vendor takes care of portability. However, this makes the web developer dependent on the plugin vendor for new features such as echo cancellation, security updates and portability to new devices. Unlike this, the Java applet has a more open development model but is not as popular among web users.

Fortunately, new versions of Flash Player have started addressing some of these restrictions, e.g., allow access to raw microphone data, and use of native hardware for video display. Moreover, some of these restrictions are not present in the AIR (Adobe Integrated Runtime) environment, e.g., it allows UDP sockets. Unlike Flash Player, AIR applications run as standalone applications on a PC but cannot run within the browser.

There are alternatives such as the Microsoft Silverlight plugin, or standards, e.g., HTML5. These alternatives are not yet as mature and ubiquitous, and potentially suffer from the same problem of limited connection and device restrictions.

## 3.3 Server Restrictions

Call signaling requires an asynchronous message exchange, e.g., to invite a user in a call. HTTP is traditionally a stateless request-response protocol with short-lived connections. In the past, developers have used several techniques to enable long-lived client-server connection to asynchronously receive events from a server. HTML5 uses websocket to enable asynchronous messages.

The Google App Engine (GAE) [20] is a cloud infrastructure that enables new web applications to readily use distributed data store and computing resources. Such cloud-based systems may impose restrictions on long lived connections due to a server's socket count limit. Recently GAE added support for short asynchronous messaging from server to client but it cannot be used as long lived general purpose TCP connection, e.g., for continuous media streaming.

A long lived persistent client-server TCP connection has other implications. It requires careful consideration of the scalability of server farms to handle many persistent connections. Any middlebox (firewall or NAT) has to keep persistent state for the connection. A corporate NAT with many clients behind it may run out of open ports. Moreover, the keep-alive traffic needed for persistent NAT binding further increases the network bandwidth.

### 3.4 Benefits of Flash Player Plugin

We have listed the limitations of browser plugin previously. Now, we will show what makes the Flash Player plugin a promising approach for web based audio and video communication.

The most important reason is that Flash Player is available to almost everyone with a computer and an Internet connection. It is available more than a specific browser brand. Anyone with a personal computer can generally use a Flash application.

The second reason is that developers find it very easy to work on the Flash platform. The programming language of Flash Player is similar to what millions of web developers are familiar with. For instances, similar to HTML for layout and Javascript for programming on the web, Flash developers use MXML for layout and Actionscript for programming on Flash Player. Due to the similarity, Javascript and Actionscript are considered as dialects of ECMAscript standard [18]. The learning curve for web developers is quick, the integrated development environment (IDE) is very friendly, and the community support is excellent.

Thirdly, an end user of a Flash-based rich Internet application experiences an immersive environment. For example, a user can chat with his or her online Facebook friends within a page using a Flash-based face-talk application as described in Section 6.4.

Fourthly, the cross browser and cross platform support of Flash Player is amazing. This makes it independent of browser wars and cross-browser incompatibility. This helps the developer to write once and run anywhere instead of being forced to use custom kludges for individual browsers. Unlike, other VoIP systems that require session negotiation and standard-compliance, interoperability is not an issue if both ends are Flash Player.

For most users, the Flash Player is already there in the browser with no additional installation. Google Chrome now comes pre-installed with Flash Player. Not having to do yet another installation is a huge plus for many scenarios, e.g, you can do Flash-based video call in cyber-café or secure machines, where you do not have permission to install and configure standalone video conferencing software such as Skype.

Due to these reasons there have been several Flash-based video communication web sites developed in recent years. It is very easy and quick for one to get started, both from the developer and user point of view. If the system just works without much installation or configuration then the end users are happy which makes them use the system again and again. If the system is quick and simple to build, then the developers are happy and motivated to add more functions, learn quickly and innovate. This allows a software project to quickly move from idea to production instead of having to make big investment up front.

The Flash Player approach works with little effort because all the complexity is hidden in the plugin. For instance, unlike a SIP-based call, there is no explicit session negotiation for matching codecs, and interoperability is non-issue if both sides use Flash Player. We focus on this approach for the rest of the paper.

## 4. Background on the Flash Player

Let us review how the Flash Player enables audio and video communications. It is built using the familiar client-server architecture. The client Flash application runs in the browser in the Flash Player sandbox. The application can connect to backend servers including web servers, media servers or third-party servers via TCP. Fig. 4 shows two browser instances displaying web pages that embed a Flash application. The Flash application uses a web server for metadata about the users and session, a signaling server to exchange asynchronous call events, and media servers to exchange real-time audio and video. Instead of a centralized media server, it could use a rendezvous service to create end-to-end media path as we describe in this section.

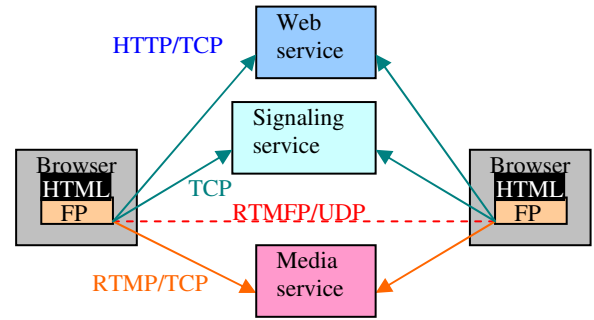

**Figure 4. Typical Flash based video communication**

### 4.1 Programming Languages

Actionscript is similar to Javascript but allows using strict type checking and modularization to build maintainable software. It is an event-driven object oriented programming language with a clean syntax. A script is compiled into a portable byte coded Flash application with the .swf extension and included in an HTML web page using standard object or embed elements. The browser invokes the Flash Player plugin to run the file.

Flex is an SDK framework for building user interface applications using markup and script. The MXML markup and ActionScript code work together to build the complete Flash application.

### 4.2 Screen Size of Application

A Flash application occupies screen space in the web page. The dimension is configured using the object attributes. By default the camera captures using an aspect ratio of 4:3. The minimum dimension must be 215x138 pixels as explained below. Flash Player displays a security prompt to the user when a Flash application tries to access camera or microphone devices. This prevents Flash Player from misuse, e.g., by automatically starting capture in application. The security prompt requires the minimum dimension of 215x138 pixels to display within the screen space of the Flash application. The device access is disabled if the application is smaller than this.

### 4.3 Elements of Video Communications

Suppose Alice and Bob want to do a video call. The Flash media service provides named streams which Alice can publish and Bob can play, and vice-versa, as shown in Fig.5. A stream represents a media flow from one publisher to zero or more players. Many server implementations are available, e.g., the commercial Flash Media Server or Wowza, or the open source Red5 or rtmplite. Web pages use an external mechanism for signaling, i.e., to exchange the stream names, e.g., putting them on a separate web page or exchanging in a separate asynchronous communication channel.

The Flex framework has four abstractions related to audio and video communication: connection, stream, devices and display. A NetConnection object represents the client-service association. It is required to create a named stream represented by NetStream. The Camera and Microphone objects provide a platform independent device input. A Flash application connects a device object to either a media stream object to send captured media to remote or a display to playback the captured media locally. The Video object enables playback of audio and video, either locally or from a NetStream. Typically, an application attaches the camera to a local video as well as published stream, and plays a stream to another video. Our Flash VideoIO project combines these abstractions into one easy to use Flash application with an extensive JavaScript API (Section 5).

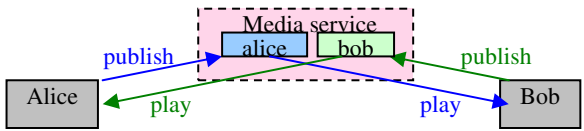

**Figure 5. Streams for two-party call**

This media-over-TCP approach works for the one-to-many streaming scenario where a provider can install multiple servers and do load sharing. It also works well for NAT and firewall traversal because of the client-server nature of the TCP connection, and since media and signaling are on the same connection. The media path latency can sometimes become large due to TCP retransmissions for reliability, and is not desirable for interactive video calls.

### 4.4 End-to-End Media Path
Recently, Flash Player added support for proprietary RTMFP to enable an end-to-end UDP media path between different plugin instances using a rendezvous service. Fig. 6 shows an example where Alice is publishing her stream and both Bob and Carol are playing it. To prevent unauthorized access of the named stream, the player also supplies the secure token of the publisher. The publisher gets a new secure token each time it connects to the rendezvous service.

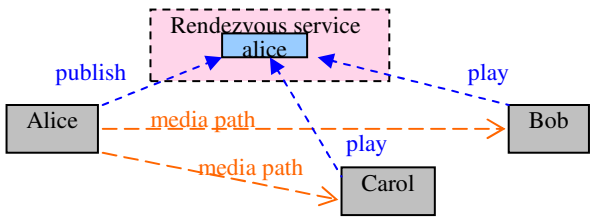

**Figure 6. RTMFP-based end-to-end media path**

As before, the web page uses an external signaling mechanism to distribute stream names and secure tokens to potential players. Both rendezvous and media traffic go over UDP. This enables low latency communication among small number of participants, but this approach does not scale if there are too many players because of the bandwidth limit on Alice's outbound link.

Adobe hosts a proprietary rendezvous service, Stratus, and provides it free for web developers. It requires a developer key, and the stream is scoped within that developer's application.

### 4.5 Group Communication
The new Flash Player has group communication capabilities using application level multicast using the rendezvous service to create a multicast tree among Flash Player instances. The one-to-many media path is distributed using multicast for more efficient group communication. A separate multicast tree originates at each publisher but may overlap with other publisher's tree. This can be used for audio, video and data distribution to many users.

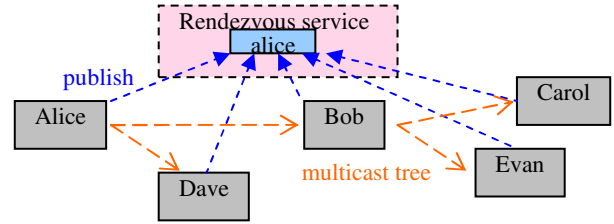

**Figure 7. RTMFP-based group communication**

Adobe hosts the multicast rendezvous service, Cirrus, and provides it free to web developers for non-commercial use. As before, the stream is scoped within the developer's key. Fig. 7 shows five participants using the rendezvous service to build the multicast tree for Alice.

### 4.6 Issues with RTMFP
Even though this UDP-based protocol is desirable for real-time video calls, it has two main problems: (1) it is proprietary and prevents interoperability with existing SIP devices, although people have started to reverse engineer the protocol, e.g., the openRTMFP project, and (2) it cannot always provide end-to-end media path because of certain firewall and NAT restrictions. In particular, if a participant is behind a UDP-blocking firewall or both participants are behind address and port dependent mapping NATs, an end-to-end media path cannot be established. Since the protocol is proprietary, a third-party cannot build scalable media relays on end-points or super-nodes similar to the Skype architecture. For a robust video conferencing service, an application provider has to invest in infrastructure to fall back to media server if needed.

## 5. API DESIGN
We have shown that implementing an audio and video call involves connecting to a media or rendezvous service and publishing/playing named streams. Connecting to the service requires an "rtmp" or "rtmfp" URL. We can specify the stream name as a URL parameter, e.g., publish=alice, so that a single URL can control the behavior of video application. This section describes the API of our Flash VideoIO project and uses its src property as the controlling URL.

The project implements a generic reusable Flash application, VideoIO.swf, with an extensive Javascript API that has the properties, methods and callbacks as summarized in Table 1. Some read-write properties control the behavior and other read-only ones indicate the status. Some properties are similar to that of the video element in HTML5, e.g., poster, autoplay, loop and controls. Others are new and used for device capture, display, connection feedback, etc. Most of the audio, video and general properties shown in the table are mapped to corresponding properties of an underlying connection, stream, devices, and display objects of Flash Player. The new properties are: url, publish, play, record, live, playing, recording, bidirection and group.

Additionally, a web application can invoke methods in VideoIO, e.g., call for RPC on the server, and receive callback events, e.g., when a property changes.

Table 1. Summary of Flash VideoIO API

| Basic properties | |
|---|---|
| src | Controls the content played or published |
| poster | Displays initial image before play or publish |
| autoplay | Whether to automatically start when src is set |
| loop | Whether to loop playback when finished |
| controls | Whether to display video control buttons |
| url | The url part of src for RTMP connection |
| publish | The publish parameter of src is stream name |
| play | The play parameter of src is stream name |
| record | Whether to record published media at server |
| live | Whether to display local camera view |
| playing | Whether to play or pause media |
| recording | Whether to record a stream |
| **Audio related properties** | |
| microphone | Whether microphone is on or not |
| codec | Audio codec: "Speex" or "Nellymoser" |
| rate | Sampling rate for Speex: 8 or 16 (kHz) |
| encodeQuality | Speex encode quality: 0 (poor) to 10 (best) |
| framesPerPacket | Speex frames per packet: 1 or 2 |
| gain | Microphone gain: 0.0 (mute) to 1.0 (max) |
| level | Current microphone capture level: 0.0 to 1.0 |
| echoSuppression | Whether echo suppression is on or not |
| silenceLevel | Microphone level of silence: 0 to 100 |
| sound | Whether sound from speaker is on or not |
| volume | Play sound volume: 0 (mute) to 1.0 (max) |
| **Video related properties** | |
| camera | Whether camera capture is on or not |
| cameraLoopback | Whether local camera view is compressed |
| cameraWidth | Width of camera capture dimension |
| cameraHeight | Height of camera capture dimension |
| cameraFPS | Desired frames/sec for camera capture |
| keyFrameInterval | Desired I-frame interval for camera capture |
| cameraQuality | Desired frame quality of encoded video |
| cameraBandwidth | Desired bandwidth of encoded video |
| videoWidth | Width in pixels of video display |
| videoHeight | Height in pixels of video display |
| currentFPS | Current frames/sec obtained from camera |
| smoothing | Whether to smooth the video display |
| fullscreen | Whether the application is in full screen |
| **General properties** | |
| currentTime | Current play or record head position in sec |
| duration | Total duration of play stream in sec |
| bytesLoaded | Total bytes loaded for play so far |
| bytesTotal | Total bytes available in play stream |
| quality | Quality of play: 0 (poor) to 1(best) |
| bandwidth | Bandwidth consumed in bytes/sec |
| bidirection | Whether to allow both publish and play |
| group | The group name to join for multicast |
| nearID | The secure token of local publisher |
| farID | The secure token of remote publisher to play |
| **Functions from Javascript to VideoIO** | |
| call | Invoke an RPC method on the server |
| postNotice | Post a text message to multicast group |
| **Event callbacks from VideoIO to Javascript** | |
| onCreationComplete | Indicates that this object is created |
| onPropertyChange | Indicates property change in this object |
| onCallback | When the server invokes RPC on client |

A web developer can either statically set a property when embedding VideoIO.swf in HTML or dynamically using Javascript.

## 5.1 The src Property

The connection URL, stream name, and other attributes are controlled by the most important property, src. You can set it to an "http" or "https" URL for playing web video files similar to the video element of HTML5. But you can play only those video formats that are supported by Flash Player. Alternatively, the src can be set for media streaming: publishing or playing, e.g., example setting the src property to rtmp://server/app?publish=alice, causes the object to connect to the RTMP server with URL rtmp://server/app and publish the locally captured audio and video to stream alice. When you set src to an rtmfp URL for publish, the object gets a nearID property after the connection to the rendezvous service is complete. This property represents the secure publisher token for the end-to-end media path. The web application should send the token to other players. The player sets it as the farID parameter in the src URL to enable a secure end-to-end path from the publisher.

## 5.2 Redundancy and Failover

If UDP transport is not feasible due to firewall/NAT restrictions, the application should fall back to client-server TCP transport, i.e., if RTMFP fails, use RTMP. To support stream failover, we plan to add a property named sources in our API to represent a list of src URLs. A publisher will attempt to publish to all the URLs in sources with at least one player, and a player will play the first URL that is successful. This has many benefits as follows:

1) *Failover of RTMFP*: In a video call, set sources to a list of rtmfp (primary) and rtmp (backup) URL in publish and play objects. A player switches to RTMP if it cannot play RTMFP. A publisher starts publishing to backup RTMP if it detects a player for that.

2) *Conference recording*: To record a conference, each publisher sets sources to a list of (1) rtmfp URL for other players and (2) rtmp URL with record=true to record the stream at the media server.

3) *Panel Discussion*: In a panel discussion, the panelists publish to a list of (1) rtmfp URL for end-to-end real-time interactive conference with other panelists, and (2) rtmfp URL with group parameter for passive viewers who receive stream over multicast.

4) *Server Failover*: A web VoIP phone uses two different gateway URLs so that it can failover to a secondary if the primary is down.

### 5.3 Use in Another Flash Application

Embedding VideoIO in a web page allows it to use the Javascript API described earlier. Sometimes, a web developer wants to use VideoIO in another Flash application instead of a web page, e.g., for consistent user experience, or picture-in-picture display of video call. All the properties, methods and callbacks described earlier are also available when VideoIO is embedded in another Flash application as a child loaded via SWFLoader in Actionscript.

The API described here is enough to implement a range of web video communication scenarios as we show next.

## 6. EXAMPLE APPLICATIONS

This section describes implementations of several common video communication scenarios, e.g., live camera view, recording and playback of multimedia messages, live video call and conferences using client-server as well as end-to-end media path technology, and broadcast of live events and panel discussion using multicast. These implementations are linked from our project page [10].

### 6.1 Live Camera View

Showing a live camera view involves setting the live property to true. The local camera view is flipped horizontally to appear as if you are looking in a mirror. This affects only the view, but not the actual media stream. We feel that mirrored local video gives the most natural experience to users in a video chat.

### 6.2 Recording and Playing Video Messages

A video messaging application requires an RTMP media server which can store recorded media files as shown in Fig. 8. When Alice wants to send a video message to Bob, she picks a stream name, say bob/msg2, to publish her audio and video. Setting src to rtmp://server/app?publish=bob/msg2&record=true connects to the media server and starts publishing local audio and video to that stream. The record mode instructs the media server to store the stream in a file, e.g., bob/msg2.flv. Once recorded, the web applications sends a notification to the receiver, e.g., via an email to Bob. Bob can play the message file by setting the src to rtmp://server/msg?play=bob/msg2.

A web server can be used to display the list of user messages and allows playback of a clicked message from within the web browser. Web authentication works for authorizing the correct receiver for viewing the message. Instead of using rtmp, the receiver may use the HTTP URL of the message on the web server to play the video file. The advantage of HTTP download over RTMP streaming is that the end user can save the video file locally. The disadvantage is that it wastes bandwidth if the user decides to ignore the message after viewing first few seconds.

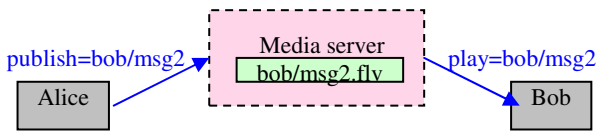

**Figure 8. Recording and playback of video messages**

The playing and recording properties control as well as indicate the current play and capture modes, respectively. When the object starts playing the stream, it sets playing to true, and when it starts recording, it sets recording to true. The web application can change these properties to start or stop playing or recording.

### 6.3 Broadcast of Live Event

The iChatNow Facebook application is a simple web application using VideoIO that allows you to publish your audio and video stream for others to view and listen. When you launch the application, it generates a random number, say 1342, as stream name, and sets the src property to rtmfp://.../?publish=1342. It uses the object's nearID, id1, and the stream name to create a play URL of the form rtmfp://.../?play=1342&farID=id1. This is used as a base64 encoded parameter in an HTTP URL and prompts you to send the URL to your friends to view your video broadcast. When your friend opens the URL in a browser, the web page extracts the play URL and sets it to src property of the embedded VideoIO object to play your published stream in full screen.

The application can be changed to use application level multicast groups as described in section 4.5. In that case, the broadcaster sets the src property with group and publish, e.g., rtmfp://.../?publish=1342&group=1342. The play URL does not include the farID, but includes the same group name, e.g., rtmfp://.../?play=1342&group=1342.

### 6.4 Two Party Video Call

The basic idea behind two-party video call was shown earlier in Fig.5. Here we combine it with an end-to-end media path provided by RTMFP. Each user has two VideoIO objects: one for publishing local media, L, and other for playing remote, R.

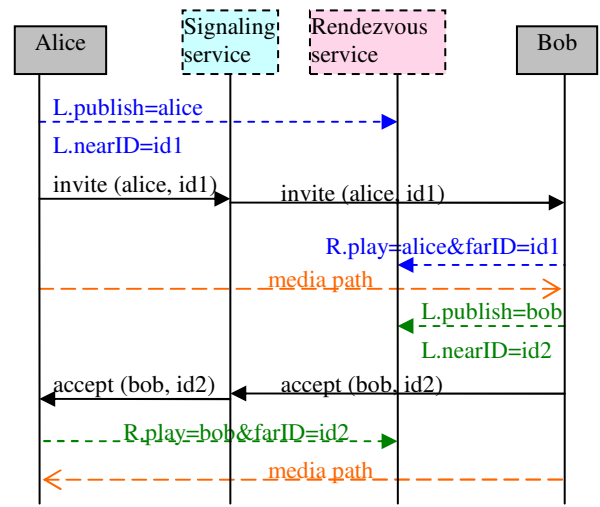

**Figure 9. Streams for two-party call**

As shown in Fig. 9, when Alice publishes her local stream using src URL of the form rtmfp://.../?publish=alice, the local object gets a new secure token in nearID=id1. Both the stream name and the secure token are sent to Bob via call signaling. Bob plays this stream on remote object using the src URL of the form rtmfp://.../?play=alice&farID=id1. Similarly, when Bob publishes his local object for stream bob, he gets a secure token, id2. Both the stream name and the secure token are sent back to Alice when Bob accepts the invitation. Alice can now play the remote object.

We need a separate *signaling* service to send some data between the two parties via call invitation and call accept. The signaling event depends on the application, but should be asynchronously delivered. Unlike Session Initiation Protocol (SIP), where a SIP proxy server does rendezvous, in VideoIO this is split into two servers because of the proprietary nature of the Flash Player. In particular, a signaling service exchanges call events with stream parameters, such as name and secure token. A rendezvous service builds an end-to-end media path or multicast tree using RTMFP. Alternatively, a media service allows a client-server media path using RTMP. We use cloud-based signaling in our examples so that our web applications are completely in the cloud.

Cloud computing and cloud-based infrastructures such as Google App Engine (GAE) have become popular among Internet consumers and small businesses because of the cost savings. Adobe offers a free rendezvous service for developers to enable an end-to-end media path and group communication. The challenge is to build a web-based video communication technology that seamlessly integrates with a cloud infrastructure so that one can quickly build applications without a large initial investment.

### 6.4.1 Random Video Chat
Random-Face is a Chatroulette-type application that randomly connects you with other people visiting the web site. When you land on the page, it prompts you for a nickname, which is used as your publishing stream name. The backend service tries to connect you with another person who is also on the page publishing his or her stream. The call signaling and media path negotiation is same as that in Fig.9. It allows audio, video and text chat with the person you are connected to.

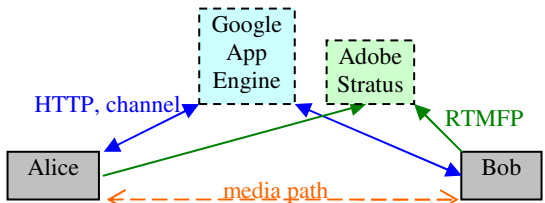

**Figure 10. Two-party random-face application**

The session initiation, discovery of other users, and messaging is done using Google App Engine (GAE)'s asynchronous channel API, whereas media stream rendezvous is done using Stratus as shown in Fig.10. The complete application is written in about 400 lines of source code using Javascript and Python.

### 6.4.2 Video Office using Google Talk
Video Office is a web-based application that allows others to visit your web-based office to talk to you. The architecture is similar to the previous application using the channel API for asynchronous messaging. The difference is that this application uses Google authentication to authenticate the office owner and uses Google chat API to notify him of the office visitors. When someone visits your video office or leaves a text message, while you are not logged in to your office, the system sends you a Google chat message indication. Additionally, the office owner sees a list of waiting office visitors, and can select a visitor to talk to, instead of automatically getting connected to the first visitor.

### 6.4.3 Video Chat with Experts
This is an extension of the previous project. It allows you to signup as an expert on some topic and to potentially monetize your time giving advice. The visitors can search for an expert based on a topic, see his calendar, sign up to talk to him, and video chat with him in real-time similar to a Video Office.

### 6.4.4 Social Video Chat
Facebook is a popular social networking platform. The Face-Talk application shows how to use VideoIO on Facebook to enable live video chat with your online friends. The system architecture is same as Fig. 9. Instead of channel API for signaling, it uses the *live messaging* and text chat available with the Facebook API.

Once you install and launch the application on your Facebook page, it publishes your local stream. It prompts you to select from currently online Facebook friends. Once selected it sends a web URL to the friend using the Facebook's text chat. The web URL encodes the necessary parameters to join your call. When the receiver clicks on the web URL, the application sends a live message containing joining parameters to the caller using the Facebook API. When both sides have the necessary parameters, they can play each other's published stream in a two-party call. When either side terminates the call, a live message notifies the other party.

## 6.5 Multi-party Video Conference
A multi-party video conference is a natural extension of the two-party video call described previously. With N participants in the conference, each participant deals with one publish stream and N-1 play streams. Thus, the web application uses N instances of VideoIO, one for publish and others for play.

The Public-Chat is a multi-party audio, video and text chat application built on top of Google App Engine and using channel API for asynchronous instant messaging and presence. It allows both public and hidden chat rooms, user listing of participants, and persistent messages using Google data store. You can publish your video stream or play the streams of others who are publishing, by clicking the checkbox next to your or other participant's name. This is also a complete cloud-based video conference service.

The Voice and Video on the Web project [19] uses a resource oriented signaling API over websocket to enable multiparty conferencing with audio, video, text chat and slide presentation. As an intermediate step to the separate application approach, it uses our Flash application to facilitate end-to-end media path. The back end is written in PHP and uses MySQL database for resource storage.

## 6.6 Broadcast of Panel Discussion
As mentioned in section 5.2, broadcasting a panel discussion is non-trivial: you need to have (1) multi-party video conference with low latency, end-to-end media path among the panelists, and (2) for bandwidth efficiency, a multicast tree from each panelist to the many viewers. Fig.11 shows three panelists, P1 to P3, and four passive viewers, V1 to V4. There is full-mesh media among the panelists, and three application level multicast trees, one from each panelist. Each panelist's media stream is color coded in the diagram as shown in the legend.

Each panelist is connected to the rendezvous service as well as media service, whereas a viewer is connected to the rendezvous service. The rendezvous service builds the multicast tree as well as the end-to-end media path. Additionally, each node connects to a separate signaling service to learn about all the panelists' streams. The complete application can be built in the cloud.

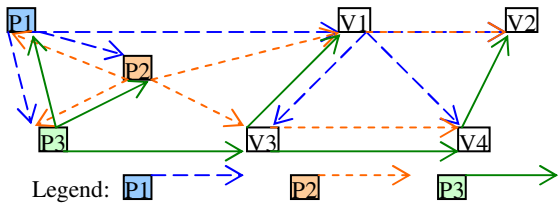

**Figure 11. Broadcast and recording a panel discussion**

The proposed sources property allows us to publish the panelist's media stream both to other panelists as well as to a multicast group for passive viewers. Additionally, the recording of panel discussions can be done by publishing to a third stream with a record parameter to a media server. Moreover, to support users behind firewalls and restrictive NATs, nodes can fall back to the media server to receive the client-server stream of the panelists.

The examples described in this section cover a range of video communication scenarios. In the next section we describe another crucial use case of how to interoperate with existing SIP systems.

## 7. INTEROPERABILITY WITH SIP/RTP

This section describes how to interoperate between a Flash application and SIP so that one can use a web browser with our VideoIO to communicate with standard SIP devices. As described earlier, the Flash Player can use RTMP or RTMFP. While RTMFP is a proprietary protocol, RTMP is now open. Many open source implementations existed even before the protocol was made open.

### 7.1 Architecture Options

Our SIP-RTMP gateway implements the translation of signaling as well as media between RTMP and SIP/RTP to support audio, video and text interoperability with SIP systems as shown in Fig. 2(a). The client side Actionscript API allows any third-party to build a user interface for the web-based phone. There are several modes in which the gateway can be used as shown in Fig. 12: (a) hosted by a telephony service provider to allow web users to make calls within that provider's system, (b) hosted by a video conferencing provider to enable phone participants, (c) running as a local application on the user's computer to turn his web browser in to a SIP phone, or (d) an independent gateway service that allows any web user to connect with any SIP network.

Due to business reasons, most current deployments follow (a) or (b) to extend an existing service. Such managed gateways are relatively simple to implement compared to general purpose gateway deployments of (c) and (d). For example, the API to initiate or receive call just needs to work with one provider, and the SIP translation has to interoperate with one system. On the other hand, the gateway running on the local host has to deal with NAT/firewall traversal as well as interoperate with many popular SIP systems. Our gateway is designed to fit any mode.

When implementing a SIP-RTMP gateway, there are three design alternatives: (1) implement the gateway as a dedicated server, (2) add SIP extension to an existing RTMP server, or (3) add RTMP extension to an existing SIP system. A dedicated server is easier to manage and configure. On the other hand other alternatives are more extensible because the same server can do two tasks. Our gateway follows alternative (2).

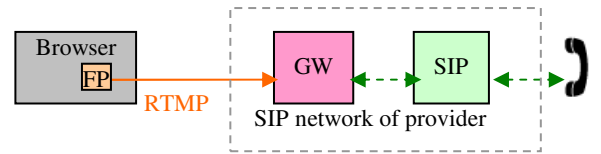

**(a) Gateway controlled by VoIP provider**

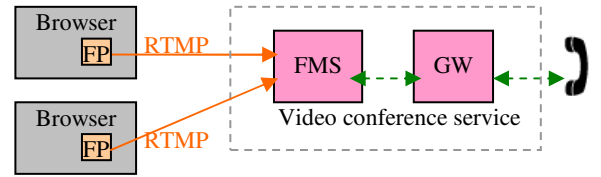

**(b) Gateway controlled by video service**

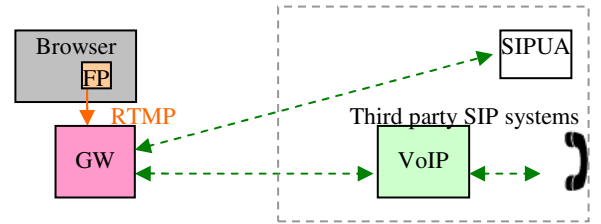

**(c) Gateway running on user's host**

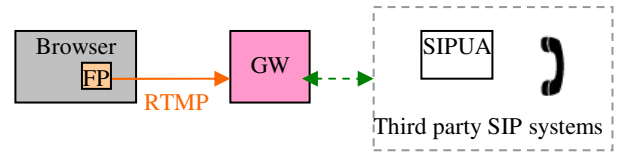

**(d) Gateway as independent service**

FMS: flash media server, GW: SIP-RTMP gateway, FP: Flash Player

**Figure 12. SIP-RTMP architecture options**

### 7.2 Client-Gateway API Options

The Flash application can connect to the gateway and invoke RPC calls. We consider the following API options.

1) *Full RPC on any connection*: The client-gateway connection is used as an RPC channel to control the SIP/RTP library running on the gateway. Every API call supplies the full context. The Flash application connects to generic gateway URL rtmp://server/sip. The application then uses RPC commands and indications to register, invite, accept or reject a SIP session. Each command has a full set of arguments needed to execute that command in the SIP library, e.g., call('invite',..., 'cid1', 'sip:alice@home.com','sip:bob@office.com') makes a call from alice to user bob in call context cid1.

The main problem is that it lacks information hiding or abstraction. Any application can alter the state of any call, unless proper authentication is implemented.

2) *A connection represents a SIP user agent*: There is one-to-one mapping between a client-gateway connection and a logical SIP user agent running on the gateway. The application connects to the gateway identifying the registering user, e.g., using URL of the form rtmp://server/sip/alice@example.net. The application also sets the registration attributes such as display name, authentication username and password. The gateway associates this connection with the user's address-of-record (AoR) alice@example.net.

Fig. 13 shows the call flows for user registration, outbound and incoming call. Some messages are not shown for brevity, e.g., SIP "180 Ringing" and ACK. The gateway sends a SIP REGISTER on behalf of connecting user, performs authentication, and keeps refreshing the registration as long as the client-gateway connection is up. The connection represents a logical SIP user agent on the gateway for this user. The application uses an RPC to invite, accept or reject a SIP session on this logical user agent.

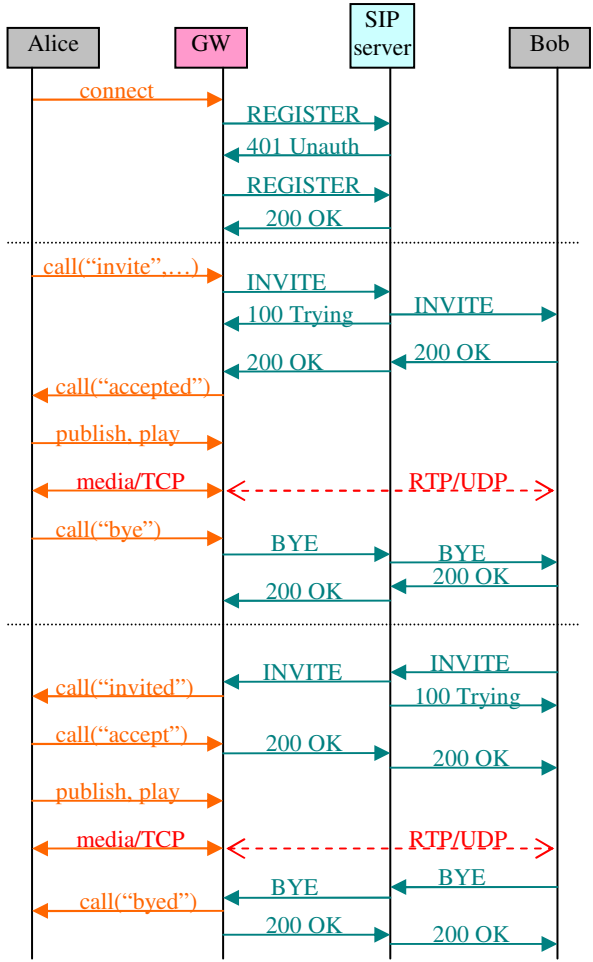

**Figure 13. Example of SIP-RTMP message flow**

To keep it simple, one user agent can be in at most one SIP session at a time. When the application invokes call('invite', ..., 'bob@home.com') the gateway sends a SIP INVITE request to the AoR sip:bob@home.com. The gateway translates between RTMP media stream of the Flash application and the RTP/UDP of the SIP side. When a call is established, the Flash application publishes to stream "local" which is consumed by the gateway and sent to SIP side. The gateway receives RTP media and publishes to stream "remote" which is consumed by the Flash application.

The call method of the VideoIO API invokes an RPC on the gateway to initiate, answer or terminate a call, or to send touch-tone digits or instant message in a call.

Instead of supporting multiple phone lines on a single logical SIP user agent, we use multiple connections for multiple SIP user agents to keep the API simple. A multi-party call is implemented entirely in the application by having multiple connections, or by making calls to a separate conference server.

3) *A connection represents a conference participant*: There is a one-to-one mapping between a client-gateway connection's scope and a SIP multi-party conference context. The application connects to a conference URL of the form rtmp://server/sip/abc1. In this case the conference is identified by scope /abc1. Each connection to this URL creates a new conference leg representing a participant. The application issues RPC to invite or accept and indications such as invited or accepted to a conference. This approach supports multiparty conferences. Each participant publishes one and receives multiple streams. The gateway informs about changes in streams whenever the members list changes.

A conference can be distributed across multiple gateways as shown in Fig. 14 with three gateway instances, four web clients and two SIP clients. The web clients connect to one of the gateway instances, whereas a SIP client can call into or be invited to the conference. The gateways communicate with each other to exchange conference membership information and transport media packets over UDP. Unlike the previous approach, if clients connect to the closest gateway, the media path on TCP is reduced to a short hop. If UDP covers most of the media path, this can improve end-to-end latency for geographically distributed participants, e.g., between A and D, and between A and S.

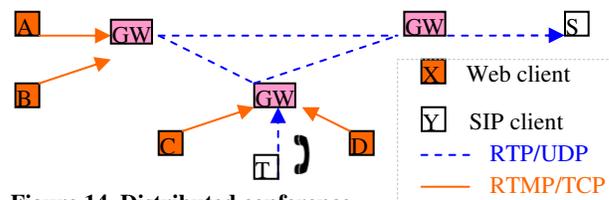

**Figure 14. Distributed conference**

This approach is more complex as it requires implementation of a multiparty conference and the consistent distribution of conference state among gateways. Ideally, we want to avoid media mixing at the gateway because: (1) not all media can be mixed easily, e.g., video, (2) audio mixing incurs an undesirable CPU processing load on the gateway, and (3) Flash Player has a built-in mixer for audio from multiple streams. Unfortunately, many existing SIP phones do not handle multiple audio streams well. This makes conferencing logic very complex in that the audio going to SIP user agents is mixed, but not to the Flash application. To keep things simple, we implemented the second approach in our gateway where one connection represents one logical SIP user agent.

### 7.3 Session Negotiation

Translation between RTMP and SIP/RTP involves both signaling and media conversion. We want to avoid media transcoding because it consumes CPU processing at the server. Although the Flash Player supports the open Speex audio codec at 8 and 16 kHz, it only supports proprietary video codec for camera captured media stream. Hence, our gateway connects the audio path and ignores video between web and SIP clients. The Speex audio codec is available in several existing SIP user agents such as X-lite as well as supported in several servers such as Asterisk. The gateway API allows selecting the sampling rate of 8 or 16 kHz in the connect call. The sampling rate of 8kHz is useful when connecting to telephony gateways which do not support 16kHz.

When the web page issues a call('invite',...) RPC to the gateway, the gateway maps it to a SIP INVITE request to the destination AoR. The session advertised in the request body using Session Description Protocol (SDP) contains the offer for both audio and video media. The audio media specifies Speex at 8 or 16 kHz, and video media specifies proprietary format named "x-flv" as shown in the following SDP fragment.

    m=audio 22700 RTP/AVP 96
    a=rtpmap:96 speex/16000
    m=video 26498 RTP/AVP 97
    a=rtpmap:97 x-flv/90000

The audio and video listening ports are randomly picked. We use dynamic RTP payload type. The "x-flv" format represents proprietary media stream of RTMP that includes both audio and video as described in Section 7.5.

If the remote side is a standard SIP user agent that does not support the advertised audio format, it responds with "488 Not Acceptable Here". If the remote side is a standard SIP user agent that supports the advertised audio format and the end user accepts the call invitation, it responds with "200 OK" and the message body contains answer SDP with only the valid audio media, i.e., the port for video media is 0. If the remote side is our gateway that understands the "x-flv" format, it responds with both audio and video media in the SDP answer. Similar SDP negotiation happens for incoming call. If incoming SDP offer does not have Speex audio codec, then we disable the audio stream. If the incoming SDP offer does not have "x-flv" video, then we disable the video stream. If none of the streams is enabled, we reject the call with "488 Not Acceptable Here". One caveat in our implementation is that the media matching is done when the Flash application accepts the incoming call. Thus, an incoming call may terminate after the web user accepts the call because of a media mismatch. This can be easily fixed in our implementation.

### 7.4 Media Transport

Standard RTP/RTCP is used for sending and receiving media packets between the gateway and the SIP user agent. The RTP timestamp is derived from the RTMP message's time field. RTMP uses millisecond whereas RTP uses clock rate for timestamp unit. For example, with 16 kHz sampling, each millisecond unit is equivalent to 16 RTP timestamps, and each Speex frame of typically 20 ms is equivalent to 320 RTP timestamps.

If the remote party supports "x-flv" video, then we assume that the remote SIP side is backed by our gateway protocol. Since "x-flv" format includes both audio and video, we do not need another Speex-based audio-only stream in the session.

### 7.5 The "x-flv" Packet Format

The "x-flv" format is not needed for interoperability with standard SIP user agent, but only for transporting Flash media between two SIP-RTMP gateways over RTP. It includes interleaved audio and video messages. Since RTMP works over reliable TCP, there is no sequence number, which makes it hard to detect and correct packet losses when translated to RTP/UDP transport. Secondly, the video message size can be huge, especially for key frames, which causes problem with RTP/UDP transport because certain NATs drop large UDP packets. For these reasons, the RTMP media message is broken down into smaller sequenced chunks such that each chunk can be sent in a single RTP/UDP packet.

The RTP timestamp is derived from the RTMP message's time field. The RTP payload type is used from SDP offer/answer. In particular for outgoing call the gateway uses payload type of 97, and for incoming call it uses the same payload type that is in offer. The sequence number, source identifier and other fields are taken care by the RTP library and are independent of the RTMP side.

The RTP payload is constructed as follows. First an RTMP media message is used in its entirety. The message header includes type, size and time attributes. These attributes are added using big-endian 32-bit number each as the header, followed by the data part of the message. Note that the data part of the RTMP message actually has one byte type information containing codec type (e.g., 0xb2 for speex/16000), but we treat the whole data part together for simplicity. Fig.15 shows the assembled media message.

Typically an audio message is already small and generates one chunk. On the other hand a large video message may generate several chunks. If the assembled message is large, it is broken down in to smaller chunks of size at most 1000 bytes. All except the last chunk will be of the same size. Typically an audio message is already small and generates one chunk. On the other hand a large video message may generate several chunks. Each chunk is treated as opaque data for subsequent formatting. Thus, the receiving side must re-assemble the full message as described above from the received chunks before processing of the message.

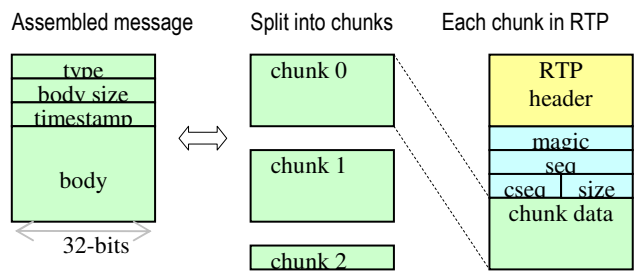

**Figure 15. "x-flv": RTMP message ⇔ RTP packet**

If the sender splits a message into chunks, the receiver must receive all the chunks to construct the whole message. Even if a single chunk is lost, the whole message needs to be discarded. Moreover, for a video message, if one message is discarded, all subsequent messages are discarded until the next key frame is received. The type byte in the data part indicates whether a message contains a key or non-key frame. This is because for encoded video from Flash Player, if part of a video frame is missing, you cannot correctly decode and render the frame until the next key frame refreshes the display. The RTMP specification also defines chunks but lacks proper sequence numbers to detect packet loss; hence our chunk algorithm is different.

Each chunk is prepended with a chunk header to form the complete RTP payload. Each chunk header starts with four bytes of the magic word 'RTMP', which is the big-endian 32-bit number 0x52544d50, to allow detecting corrupted or incorrectly received data. There are two sequence numbers: the message sequence number (seq) and chunk sequence number (cseq). Each assembled message gets a unique auto-incremented seq number. The first chunk of a message has cseq of 0. If a message is broken into, say, three chunks, then the chunks will get cseq as 0 to 2 in that order. In the chunk header, next 32-bits contain the big-endian message sequence number, seq. The RTP sequence number is based on the lower layer's actual message sent count, whereas seq is based on

RTMP's message count. A single RTMP message may be split in to multiple chunks, hence multiple RTP packets. The `seq` is followed by a big-endian 16-bit chunk sequence number, `cseq`. The next 16-bit field is an optional size of the assembled message and is present only for the first chunk of the message. This gives the full size of the assembled message for receiver to know when to re-assemble. Alternatively, the body size could be used, but makes the code more involved at the receiver. The chunking algorithm makes sure that each chunk including the chunk header and RTP header is small enough to fit within 1500 bytes. The sender is expected to send all the messages in the correct `seq` order, and all the chunks of a message in the correct `cseq` order.

The header overhead is significant, especially for audio payload, with 12 bytes of RTP, 12 bytes of "x-flv" and 12-bytes of assembled message header. Note however that "x-flv" is useful only between two gateways, and not used for interoperating with standard SIP phones. Secondly, the "x-flv" header size can be reduced to only 4 bytes by removing the magic work, reducing seq to 2 bytes similar to RTP sequence number, and using size from assembled message header. The assembled message header can be reduced to 3 bytes by reducing type to 1 byte, body size to 2 bytes and re-using RTP timestamp. We will incorporate these optimizations if inter-gateway call becomes popular in practice.

The receiver processing is described below. First the payload type is matched to identify it as an "x-flv" packet as per earlier SDP negotiation. The other RTP fields such as timestamp can be ignored because they also appear in the actual assembled message. The payload of the RTP packet is parsed using the chunk format described above. The receiver verifies the magic word of 'RTMP', and ignores the packet otherwise. The message sequence number is extracted as `seq`. If the `seq` is 0, then the message `size` is extracted. Remaining data is assumed to be chunk data. The receiver maintains the last `seq` received so far, and all the chunk data in the last `seq` received so far. The receiver may maintain more than one `seq` data, if it wants to handle out-of-order packets. For each received packet, the receiver checks if all the chunks are received or not? If the total size of the entire chunk data received so far becomes equal to the message size found in the first chunk, then we have received all the chunks. When all the chunks are received, the entire chunk data is concatenated using `cseq` in order to create a complete assembled message. This message is than converted to an RTMP message and sent to the Flash application.

Note that the receiver may detect lost chunks if there is a missing `cseq` and will discard all the chunks in that message `seq`. The receiver may also detect missing a first chunk if the new `seq` number is more than the last `seq` but the new `cseq` is not 0. In this case it will discard all future chunks in this `seq`.

## 8. CONCLUSIONS AND FUTURE WORK

We have described the various challenges of web-based audio and video communication, and presented in detail a generic API using our Flash application. We showed how to implement several web video communication scenarios. This paper presented both the advantages and disadvantages of using the Flash Player plugin. The proposed API can be used as a baseline for future extensions to HTML for video communication.

Our `VideoIO` application has some restrictions, but presents a very high-level abstraction for developers to quickly build video applications using HTML/Javascript. The planned redundancy and failover mechanism enables many new scenarios such as conference recording and panel discussion in addition to media failover from an end-to-end UDP to a client-server TCP, if end-to-end UDP does not work due to firewall/NAT.

We also described various interoperating scenarios between web-based Flash application and standard SIP systems. Our translation mechanism allows audio interaction with third-party SIP end points as well as audio and video interaction with other Flash-enabled endpoints behind another gateway. The proposed "x-flv" payload format allows inter-gateway media over RTP.

We have measured the performance of our gateway software, and have also ported it to use efficient `gevent` extension for sockets and I/O, which reduced the CPU usage by 50%. In steady state, it takes about 66 MHz per voice call going through the gateway with 8 kHz Speex codec on Intel-based modern CPUs. Thus, on a quad-core 2 GHz modern machine, it can support over hundred simultaneous calls.

We plan to improve our open source gateway as follows. We will allow `rtmps` URLs and `sips` AoRs to perform secure signaling between the web to gateway and the gateway to SIP system. Audio transcoding is useful because Speex is unavailable in older versions of the Flash Player and some telephony gateways. Our current software supports interoperability with G.711 voice codec, but we plan to add support for additional VoIP codecs. Video transcoding between Flash Player's proprietary video codec and H.264 allows interoperability with third-party SIP system. We will add NAT and firewall traversal so that the gateway can be used on user's host computer for better performance. We also plan to port CPU intensive media processing in the gateway from Python to C/C++ for higher performance.

Both our software components, the Flash `VideoIO` and the SIP-RTMP gateway as part of a media server, `rtmplite`, are being used actively by many web developers to quickly build web-based video applications and interoperate with SIP infrastructure.

## 9. ACKNOWLEDGMENTS

We thank Alan Johnston, Henry Sinnreich and Wilhelm Wimmreuter for thorough review of this paper and insightful discussions on web video communication. Gaurav Gupta helped in implementing the Facebook application using the `VideoIO` API. Thanks to the many web developers who have used and given feedback on our Flash `VideoIO` and SIP-RTMP projects.